\documentclass[journal,10pt]{IEEEtran}
\usepackage{amsfonts}
\usepackage{amsmath,graphicx}
\usepackage{lipsum}
\usepackage{algorithm}
\usepackage{algorithmic}
\usepackage{amsmath,graphicx}
\usepackage{lipsum}
\usepackage{algorithm}
\usepackage{algorithmic}
\usepackage{subfigure}
\usepackage{setspace}
\usepackage{epstopdf}
\usepackage{bbm}
\usepackage{dsfont}

\usepackage{color,soul}

\begin{document}

\title{Experimental Design via Generalized Mean Objective Cost of Uncertainty%
}
\author{Shahin~Boluki,~ Xiaoning~Qian,~ and~Edward~R.~Dougherty\thanks{%
S. Boluki, X. Qian and E. R. Dougherty are with the Department of Electrical
and Computer Engineering, Texas A\&M University, College Station, TX 77843
USA. Email: (s.boluki@tamu.edu, xqian@ece.tamu.edu, and edward@ece.tamu.edu).%
}\vspace{-0.25in}}
\maketitle

\begin{abstract}
The mean objective cost of uncertainty (MOCU) quantifies the performance
cost of using an operator that is optimal across an uncertainty class of
systems as opposed to using an operator that is optimal for a particular
system. MOCU-based experimental design selects an experiment to maximally
reduce MOCU, thereby gaining the greatest reduction of uncertainty impacting
the operational objective. The original formulation applied to finding
optimal system operators, where optimality is with respect to a cost
function, such as mean-square error; and the prior distribution governing
the uncertainty class relates directly to the underlying physical system.
Here we provide a generalized MOCU and the corresponding experimental
design. We then demonstrate how this new formulation includes as special
cases MOCU-based experimental design methods developed for materials science
and genomic networks when there is experimental error. Most importantly, we
show that the classical Knowledge Gradient and Efficient Global Optimization
experimental design procedures are actually implementations of MOCU-based
experimental design under their modeling assumptions.
\end{abstract}

%\author{Shahin~Boluki$^{*}$,
%%~\IEEEmembership{Student~Member,~IEEE,}
%Xiaoning~Qian, 
%%\IEEEmembership{Member,~IEEE,} 
%and~Edward~R.~Dougherty
%%\IEEEmembership{Fellow,~IEEE} 
%\thanks{
%S. Boluki, X. Qian and E. R. Dougherty are with the Department of Electrical
%and Computer Engineering, Texas A\&M University, College Station, TX 77843
%USA. Email: (s.boluki@tamu.edu, xqian@ece.tamu.edu, and edward@ece.tamu.edu).}
%\thanks{$^{*}$ Corresponding author.}}

%\ninept
%
\allowdisplaybreaks[1]

%\vspace{-0.05in}
\section{Introduction}
%\vspace{-0.05in}

The mean objective cost of uncertainty (MOCU) quantifies the performance
cost of using an operator that is optimal across an uncertainty class of
systems as opposed to an operator that is optimal for a particular system
within the class \cite{mocu}. MOCU-based experimental design selects an
experiment that maximally reduces MOCU, thereby optimally reducing
uncertainty with respect to the operational objective \cite%
{Dehghannasirimocu}. For instance, if one wishes to design a Wiener filter
when the relevant power spectra not fully known but belong to an uncertainty
class of power spectra, then the problem is to design a linear filter that
is optimal relative to both mean-square error (MSE) and the probability mass
over the uncertainty class. An optimal experiment maximally reduces MOCU
relative to uncertainty in the relevant power spectra \cite{roozbehCanonical}%
.

This letter provides a generalized formulation of MOCU not necessarily
dependent on the particularities of the underlying system model or involving
a design problem focused on operators. We show that the corresponding
generalized experimental design encompasses existing formulations in signal
processing, genomics, and materials discovery, and that it fits within
Lindley's paradigm for Bayesian experimental design~\cite{ldvbsar72sa}. Within this generalized framework we examine the connection and differences of MOCU-based
formulations with other Bayesian experimental design methods. In particular,
we show that the generalized MOCU generates the same policies as Knowledge
Gradient (KG) \cite{kg2008,correlatedKG} and Efficient Global Optimization~(EGO) \cite{ego} under their modeling
assumptions, that is, for optimal experimental design under Gaussian belief
and observation noise for an offline ranking and selection problem. Not only
does the generalized MOCU framework unify disparate problems, it opens up
Bayesian experimental design for reduction of objective related uncertainty, as
demonstrated by materials discovery using Ginzburg-Landau theory.

\section{Generalized MOCU}
\label{sec:GeneralMOCU}
%\vspace{-0.05in}

We first formulate experimental design in terms of generalized MOCU and then
give the standard method by simply defining the terms in the generalized
model appropriately. {In this letter, the lower case Greek letters denote random variables or distribution functions and capital Greek letters denote the corresponding domain space.} We assume a probability space $\Theta $ with
probability measure $\pi $, a set $\Psi $, and a function $C:\Theta \times
\Psi \rightarrow \lbrack 0,\infty )$, where $\Theta ,\pi ,\Psi $, and $C$
are called the \emph{uncertainty class}, \emph{prior distribution}, \emph{%
action space}, and \emph{cost function}, respectively. Elements of $\Theta $
and $\Psi $ are called \emph{uncertainty parameters} and \emph{actions},
respectively. For any $\theta \in \Theta $, an \emph{optimal action} is an
element $\psi _{\theta }\in \Psi $ such that $C(\theta ,\psi _{\theta })\leq
C(\theta ,\psi )$ for any $\psi \in \Psi $. An \emph{intrinsically Bayesian
robust (IBR) action} is an element $\psi _{\mathrm{IBR}}^{\Theta }\in \Psi $
such that $\mathrm{E}_{\theta }[C(\theta ,\psi _{\mathrm{IBR}}^{\Theta
})]\leq \mathrm{E}_{\theta }[C(\theta ,\psi )]$\ for any $\psi \in \Psi $.

Whereas $\psi _{\mathrm{IBR}}^{\Theta }$ is optimal over $\Theta $, for $%
\theta \in \Theta $, $\psi _{\theta }$ is optimal relative to $\theta $. The 
\emph{objective cost of uncertainty} is defined by the performance loss of
applying $\psi _{\mathrm{IBR}}^{\Theta }$ instead of $\psi _{\theta }$ on $%
\theta $:
\begin{equation}
\mathrm{U}_{\Psi }(\Theta )=C(\theta ,\psi _{\mathrm{IBR}}^{\Theta
})-C(\theta ,\psi _{\theta }).  \label{OCU}
\end{equation}%
Averaging this cost over $\Theta $ gives the \emph{mean objective cost of
uncertainty (MOCU)}:
\begin{equation}
\mathrm{M}_{\Psi }(\Theta )=\mathrm{E}_{\theta }[C(\theta ,\psi _{\mathrm{IBR%
}}^{\Theta })-C(\theta ,\psi _{\theta })].  \label{MOCU}
\end{equation}%
The action space is arbitrary so long as the cost function is defined on $%
\Theta \times \Psi $. It can be a set of filters defined on a random process
with $C$ being mean-square error or a set of drug interventions with $C$
quantifying patient condition.

% Yoon et al. (2013)

As noted in~\cite{mocu}, MOCU can be viewed as the minimum expected value of a Bayesian loss function that maps an operator to its differential cost (for using the given operator instead of an optimal operator). The minimum expectation is attained by an optimal robust operator that minimizes the average differential cost. In decision theory, this differential cost is called the \emph{regret}, which is defined as the difference between the maximum payoff (for making an optimal decision) and the actual payoff (for the decision that has been made). From this perspective, MOCU can be viewed as the minimum expected regret for using a robust operator.

Suppose there is a set $\Xi $, called the \emph{experiment space}, whose
elements, $\xi $, called \emph{experiments}, are jointly distributed with
the uncertainty parameters $\theta $. Given $\xi \in \Xi $, the conditional
distribution $\pi (\theta |\xi )$ is the \emph{posterior distribution}
relative to $\xi $ and $\Theta |\xi $ denotes the corresponding probability
space, called the \emph{conditional uncertainty class}. Relative to $\Theta
|\xi $, we define IBR actions $\psi _{\mathrm{IBR}}^{\Theta |\xi }$ and the
conditional (remaining) MOCU, 
\begin{equation}
\mathrm{M}_{\Psi }(\Theta |\xi )=\mathrm{E}_{\theta |\xi }[C(\theta ,\psi _{%
\mathrm{IBR}}^{\Theta |\xi })-C(\theta ,\psi _{\theta })],
\label{remaining_MOCU}
\end{equation}%
where the expectation is with respect to $\pi (\theta |\xi )$. Taking the
expectation over $\xi $ gives the expected remaining MOCU, 
\begin{equation}
\mathrm{D}_{\Psi }\mathcal{(}\Theta ,\xi )=\mathrm{E}_{\mathcal{\xi }}[%
\mathrm{M}_{\Psi }(\Theta |\xi )]=\mathrm{E}_{\mathcal{\xi }}[\mathrm{E}%
_{\theta |\xi }[C(\theta ,\psi _{\mathrm{IBR}}^{\Theta |\xi })-C(\theta
,\psi _{\theta })]],  \label{eq:mocu-experimentvalue}
\end{equation}%
which is called the \emph{experimental design value}. An optimal experiment $%
\xi ^{\ast }\in \Xi $ minimizes $\mathrm{D}_{\Psi }\mathcal{(}\Theta ,\xi )$%
, i.e., 
\begin{equation}
\xi ^{\ast }=\arg \!\min_{\xi \in \Xi }\mathrm{D}_{\Psi }\mathcal{(}\Theta
,\xi ).  \label{eq:mocu-policy}
\end{equation}%
$\xi ^{\ast }$ also minimizes the difference between the expected remaining
MOCU and the current MOCU:% (maximizes the absolute difference): 
\begin{equation}
\begin{split}
\xi ^{\ast }=& \arg \!\min_{\xi \in \Xi }\mathrm{D}_{\Psi }\mathcal{(}\Theta
,\xi )-\mathrm{M}_{\Psi }(\Theta ) \\
=& \arg \!\min_{\xi \in \Xi }\mathrm{E}_{\mathcal{\xi }}[\mathrm{E}_{\theta
|\xi }[C(\theta ,\psi _{\mathrm{IBR}}^{\Theta |\xi })-C(\theta ,\psi
_{\theta })]]- \\
& \quad \quad \quad \quad \mathrm{E}_{\theta }[C(\theta ,\psi _{\mathrm{IBR}%
}^{\Theta })-C(\theta ,\psi _{\theta })] \\
=& \arg \!\min_{\xi \in \Xi }\mathrm{E}_{\mathcal{\xi }}[\mathrm{E}_{\theta
|\xi }[C(\theta ,\psi _{\mathrm{IBR}}^{\Theta |\xi })]]-\mathrm{E}_{\theta
}[C(\theta ,\psi _{\mathrm{IBR}}^{\Theta })].
\end{split}
\label{eq:mocu-policy2}
\end{equation}

With sequential experiments, the action space and experiment space can
be time dependent, i.e., they can be different for each time step.
Hereafter, in sequential experiment setups, the action space and experiment
space at time step $t$, and the optimal experiment selected at $t$ to be
performed at the next time step are denoted by $\Psi ^{t}$, $\Xi ^{t}$, and $%
\xi^{\ast ,t}$, respectively. Let $\pi (\theta |\xi ^{:t})$ be the posterior
distribution after observing the selected experiments' outcomes from the
first time step through $t$, and $\Theta |\xi ^{:t}$ denote the
corresponding conditional uncertainty class. When experiments are selected
sequentially and there is no fixed limited budget of experiments but instead
the experimenter wants to stop the iterative procedure when only negligible
knowledge regarding the objective can be gained from additional experiments,
the form in \eqref{eq:mocu-policy2} is useful because it incorporates the
difference between the expected remaining MOCU and the current MOCU. The
iterative procedure may be stopped if it falls below a threshold. While this
procedure is optimal at each step, it is not optimal given a fixed number of
experiments to be performed. This latter kind of finite-horizon optimal
design using MOCU is treated in \cite{Imani_bioinf} using dynamic
programming.

In the standard formulation, MOCU depends on a class of operators applied to
a parameterized physical model in which $\theta $ is a random vector whose
distribution depends on a physical characterization of the uncertainty. For
instance, in a gene regulatory network, uncertainty arises regarding
regulations and experimental design decides which unknown regulations should
be determined via experiments so as to minimize the cost of uncertainty
relative to the objective of minimizing the long-run likelihood of the cell
being in a cancerous state \cite{mocu,Dehghannasirimocu,Navid}. $\Theta $ is
an uncertainty class of system models parameterized by a vector $\theta $
governed by a probability distribution $\pi (\theta )$ and $\Psi $ is a
class of operators on the models whose performances are measured by $C$. For
each operator $\psi $, $C(\theta ,\psi )$ is the cost of applying $\psi $ on
model $\theta \in \Theta $. Initially proposed for optimal intervention in
Markovian regulatory networks \cite{mocu} and optimal robust classification 
\cite{DALTON1}, IBR operators have been designed for linear and
morphological filters \cite{DALTON2} and Kalman filters \cite{RoozbehKalman}.

As originally formulated \cite{Dehghannasirimocu}, experimental design
involves $k$ experiments $T_{1},\ldots ,T_{k}$, where experiment $T_{i}$
exactly determines the uncertain parameter $\theta _{i}$ in $\theta =(\theta
_{1},\theta _{2},\ldots ,\theta _{k})\in \Theta $. The \emph{conditional
uncertainty vector} $\theta |\theta _{i}$ is composed of all uncertain
parameters other than $\theta _{i}$, with $\theta _{i}$ now determined by $%
T_{i}$. $\Theta |\theta _{i}$ is the \emph{reduced uncertainty class} given $%
\theta _{i}$. The IBR operator for $\Theta |\theta _{i}$, the remaining MOCU
given $\theta _{i}$, and the experimental design value take the forms $\psi
_{\mathrm{IBR}}^{\Theta |\theta _{i}}$, $\mathrm{M}_{\Psi }(\Theta |\theta
_{i})$, and $\mathrm{D}\mathcal{(}\theta _{i})=\mathrm{E}_{\theta _{i}}[%
\mathrm{M}_{\Psi}(\Theta |\theta _{i})]$, respectively. The \emph{optimal
experiment} $T_{i^{\ast }}$ is specified by $i^{\ast }=\arg \min_{i=1,...,k}%
\mathrm{D}\mathcal{(}\theta _{i})$.

Returning to the generalized MOCU formulation, there is wide flexibility in
experimental design, depending on the assumptions regarding the uncertainty
class, action space, and experiment space, leading to many existing Bayesian
experimental design formulations. Bayesian experimental design has a long
history, in particular, utilizing the expected gain in Shannon information 
\cite{s59core59as,gh63,DeGroot1986,b79}. In 1972, Lindley
proposed a general decision theoretic approach incorporating a two-part
decision involving the selection of an experiment followed by a terminal
decision \cite{ldvbsar72sa}. Supposing $\lambda $ is a design selected from
a family $\Lambda $ and $\mathbf{X}$ is a data vector, and leaving out the
terminal decision, an optimal experiment is given by 
\begin{equation}
\lambda ^{\ast }=\arg \max_{\lambda \in \Lambda }\mathrm{E}_{\mathbf{X}}[%
\mathrm{E}_{\Theta }\left[ U(\theta ,\mathbf{X},\lambda )|\mathbf{X},\lambda %
\right] |\lambda ],  \label{Lindley}
\end{equation}%
where $U$ is a utility function (see \cite{cv95} for the full
decision-theoretic optimization).

With generalized MOCU, each experiment $\xi $ corresponds to a data vector $%
\mathbf{X}|\xi $ and the expected remaining MOCU is 
\begin{equation}
\begin{split}
&\mathrm{E}_{\xi }[\mathrm{M}_{\Psi }(\Theta |\mathbf{X},\xi )]\  \\
&=\mathrm{E}_{\mathbf{X}|\xi }[\mathrm{E}_{\Theta }[C_{\theta |(\mathbf{X}%
|\xi )}(\psi _{\mathrm{IBR}}^{\Theta |(\mathbf{X}|\xi )}) -C_{\theta |(%
\mathbf{X}|\xi )}(\psi _{\theta |(\mathbf{X}|\xi )})]] \\
& =\mathrm{E}_{\mathbf{X}|\xi }[\mathrm{E}_{\Theta }[\mathrm{U}_{\Psi
}(\theta ,\mathbf{X},\xi ;\Theta )]].
\end{split}
\label{remaining_MOCU_data}
\end{equation}%
%
%
%
%
%\begin{eqnarray}
%\mathrm{E}_{\xi }[\mathrm{M}_{\Psi }(\Theta |\mathbf{X},\xi )]\ &=&\mathrm{E}%
%_{\mathbf{X}|\xi }[\mathrm{E}_{\Theta }[C_{\theta |(\mathbf{X}|\xi )}(\psi _{%
%\mathrm{IBR}}^{\Theta |(\mathbf{X}|\xi )})  \notag \\
%&&-C_{\theta |(\mathbf{X}|\xi )}(\psi _{\theta |(\mathbf{X}|\xi )})]] \\
%&=&\mathrm{E}_{\mathbf{X}|\xi }[\mathrm{E}_{\Theta }[\mathrm{U}_{\Psi
%}(\theta ,\mathbf{X},\xi ;\Theta )]].  \label{remaining_MOCU_data}
%\end{eqnarray}%
From \eqref{remaining_MOCU_data}, the optimization of \eqref{eq:mocu-policy}
can be expressed in the same form as \eqref{Lindley}, with $\xi $ in place
of $\lambda $ and utility function $-\mathrm{U}_{\Psi }(\theta ,\mathbf{X}%
,\xi ;\Theta )$.

%in~\citep{mocu,Dehghannasirimocu}. Qian and Dougherty (2016)

Hence, in descending order of generality, we have Lindley's procedure,
generalized MOCU, and MOCU. The salient point regarding the latter is that
the uncertainty is on the underlying random process, meaning the science,
and its aim is to design a better operator on the underlying process. As stated in~\cite{Qian2016}, there is a \emph{scientific gap} in constructing functional models and making prior assumptions on model parameters when the actual uncertainty applies to the underlying random processes. We
next show how generalized MOCU includes other existing objective-based
experimental-design formulations.

\section{Guiding Simulations in Materials Discovery}
\label{sec:Materials}
%\vspace{-0.05in}
In \cite{RoozbehMaterial}, optimal experimental design based on MOCU is
applied to a computational problem for shape memory alloy~(SMA) design with
desired stress-strain profiles for a particular dopant at a given
concentration utilizing time-dependent Ginzburg-Landau (TDGL) theory. The
TDGL model simulates the free energy for a specific dopant with a specified
concentration, given the dopant's parameters. The assumption is that there
is a set $D=\{d_{1},\ldots,d_{N}\}$ of $N$ potential dopants and each dopant 
$d_{i}$ can be characterized by two parameters, its strength $h_{i}$ and its
range of stress disturbance $r_{i}$. The concentration of the dopants can be
selected from a set $O=\{o_{1},\ldots,o_{P}\}$ of $P$ pre-specified values.
The true values of these dopant parameters are unknown; however, there
exists a prior distribution over the dopant parameters. In summary, we have $%
\Theta =H\times R$ and $\theta =[h,r]$, where $h=[h_{1},\ldots,h_{N}]$ and $%
r=[r_{1},\ldots,r_{N}]$, and $H$ and $R$ represent the sample spaces of $h$
and $r$, respectively. Thus, $\theta _{i}=[h_{i},r_{i}]$ fully characterizes
dopant $d_{i}$.

Since the computational complexity of the TDGL model is enormous, the goal
is to find an optimal dopant and concentration to minimize the simulated
energy dissipation, with the least number of times running the TDGL model
(least number of experiments). Following \cite{RoozbehMaterial}, for this
purpose, a surrogate model $g(h,r,o)$ is trained based on fitting some
initial data generated from the TDGL model. The surrogate model can
approximately predict a dissipation energy for a specified dopant and
concentration, and it is used as the cost function throughout the
experimental design iterations. The TDGL model acts as the true underlying
system, or Nature, and the surrogate model is the model of the true system.
The action space is $\Psi =\{\psi _{d_{i},o_{j}}\}_{d_{i}\in D,o_{j}\in O}$,
where each action $\psi _{d_{i},o_{j}}$ is using the $i$\textsuperscript{th}
dopant with the $j$\textsuperscript{th} possible concentration. The cost
function is $C(\theta ,\psi _{d_{i},o_{j}})=g(h_{i},r_{i},o_{j})$. The
experiment space is $\Xi =\{\xi _{d_{i},o_{j}}\}_{d_{i}\in D,o_{j}\in O}$,
where $\xi _{d_{i},o_{j}}$ corresponds to obtaining a noisy measurement of
the dissipation energy when using the $i$\textsuperscript{th} dopant with
the $j$\textsuperscript{th} concentration. $\xi _{d_{i},o_{j}}\sim f(\xi
_{d_{i},o_{j}}|\theta _{i})$ , where $f$ is a probability distribution.

In this framework, the IBR action at time step $t$ is 
\begin{equation}
\begin{split}
\psi _{\mathrm{IBR}}^{\Theta |\xi ^{:t}}=& \arg \!\min_{\psi \in \Psi
}E_{\theta |\xi ^{:t}}\big[C(\theta ,\psi )\big] \\
=& \arg \!\min_{\psi _{d_{i},o_{j}}\in \Psi }E_{\theta |\xi ^{:t}}\big[%
g(h_{i},r_{i},o_{j})].
\end{split}%
\end{equation}%
From \eqref{eq:mocu-experimentvalue} and \eqref{eq:mocu-policy}, the optimal
experiment at time step $t$ is 
\begin{equation}
\begin{split}
& \xi^{\ast ,t}=\arg \!\min_{\xi \in \Xi }\mathrm{E}_{\mathcal{\xi }}[%
\mathrm{E}_{\theta |\xi }[C(\theta ,\psi _{\mathrm{IBR}}^{\Theta |\xi ,\xi
^{:t}})-C(\theta ,\psi _{\theta })]] \\
& =\arg \!\min_{\xi _{d_{i},o_{j}}\in \Xi }\mathrm{E}_{\mathcal{\xi }%
_{d_{i},o_{j}}}[\mathrm{E}_{\theta |\xi _{d_{i},o_{j}},\xi ^{:t}}[C(\theta
,\psi _{\mathrm{IBR}}^{\Theta |\xi ^{:t+1}})]],
\end{split}
\label{eq:mocu-material}
\end{equation}%
where the second equality is due to the independence of $C(\theta ,\psi
_{\theta })$ from $\xi _{d_{i},o_{j}}$. The last line of %
\eqref{eq:mocu-material} is exactly the policy proposed in \cite%
{RoozbehMaterial} for this materials science problem.

\section{Dynamical Genetic Networks }
\label{sec:Life}
In \cite{Navid}, optimal objective-based experimental design is derived for
networks with multiple dynamic trajectories, modeling in \cite{Navid} is
based on \cite{Navid1}. Briefly, the network's nodes and their corresponding
values represent entities, proteins/chemicals or genes, and their
corresponding concentration levels or expression levels, respectively. The
values are assumed to be nonnegative integers. Each edge represents an
interaction with its input, regulation, and output nodes. Each interaction
can dynamically happen if all of its input and activator nodes are nonzero
and its inhibitor nodes are zero. All interactions are known. When the
network is in state $x$, it can have one or more possible interactions based
on the node values, where if any takes place, the network transitions to a
next state. When multiple interactions exist, if knowledge of the relative
priorities of these competing interactions exist, we can completely
determine the state trajectory of the network from an initial state $x_{0}$.

The assumption is that these relative priorities are not known but can be
measured one at a time with experimental error. If the network has $R$ of
these competing interactions, i.e., interactions that can dynamically happen
at the same time, then the uncertainty class consists of a set of $R$
Boolean random variables, $\Theta =\{0,1\}^{R}$, and $\theta =(\theta
_{1},...,\theta _{R})$, where $\theta _{i}\in \{0,1\}_{i=1,...,R}$. The $i$%
\textsuperscript{th} experiment can determine the value of $\theta _{i}$
with an experimental error having probability $\delta _{i}$. Specifically,
if $\theta _{i}$ is selected to be measured, with probability $1-\delta _{i}$
the outcome of the experiment is $\theta _{i}$, and with probability $\delta
_{i}$ is $1-\theta _{i}$. Here, $\Xi =\{\xi _{1},...,\xi _{R}\}$, each
experiment $\xi _{i}$ corresponds to measuring $\theta _{i}$, and 
\begin{equation}
\xi _{i}|\theta _{i}=%
\begin{cases}
\theta _{i}\quad \text{with probability}\quad 1-\delta _{i}, \\ 
1-\theta _{i}\quad \text{with probability}\quad \delta _{i}.%
\end{cases}%
\end{equation}%
An action blocks an interaction from happening, so the action space is $\Psi
=\{\psi _{1},...,\psi _{A}\}$, where $A$ is the number of interactions that
can be blocked. Each action changes the dynamic trajectory of the network.
If the set of possible state trajectories is denoted by $S_{\psi
_{i}}^{\Theta }$ when the $i$\textsuperscript{th} action ($\psi _{i}$) is
taken, then the probability of each trajectory $s\in S_{\psi _{i}}^{\Theta }$
is 
\begin{equation}
P_{S_{\psi _{i}}^{\Theta }}(s)=\mathrm{E}_{x_{0}}\big[\mathrm{E}_{\theta }[%
\mathds{1}_{s_{x_{0},\theta }(\psi _{i})=s}]\big],
\end{equation}%
where $\mathds{1}_{w}$ is the indicator function ($\mathds{1}_{w}=1$ if $w$
is true and is 0 otherwise), and $s_{x_{0},\theta }(\psi _{i})$ is the
deterministic trajectory for a fixed initial state $x_{0}$ and $\theta $,
when action $\psi _{i}$ is taken. Here, $S_{\psi _{i}}^{\Theta }=\cup
_{x_{0}\in X_{0}}\cup _{\theta \in \Theta }s_{x_{0},\theta }(\psi _{i})$,
where $X_0$ denotes the set of all possible initial states. For each
trajectory $s$, the dynamic performance cost $\varepsilon (s)$ is defined as
the distance (in terms of any appropriate norm) of the steady-state vector
corresponding to that trajectory ($x_{f}^{s}$) from a desired distribution $%
v $, i.e. $\varepsilon (s)=||x_{f}^{s}-v||$. Thus, the cost function for a
fixed $\theta $ and action $\psi $ is the expected cost over the possible
trajectories, $C(\theta ,\psi )=\mathrm{E}_{S_{\psi }^{\Theta }}[\varepsilon
(s)]$.

The IBR action for this problem is 
\begin{equation}
\psi _{\mathrm{IBR}}^{\Theta }=\arg \!\min_{\psi \in \{\psi _{1},...,\psi
_{A}\}}\mathrm{E}_{\theta }[C(\theta ,\psi )].
\end{equation}%
According to %
\eqref{eq:mocu-experimentvalue} and \eqref{eq:mocu-policy}, the optimal
experiment can be derived as %\begin{equation}
%\begin{split}
\begin{align}
 \xi^{\ast }&= \arg \!\min_{\xi _{i}\in \Xi }\mathrm{E}_{\xi _{i}}[\mathrm{%
E}_{\theta |\xi _{i}}[C(\theta ,\psi _{\mathrm{IBR}}^{\Theta |\xi _{i}})-C(\theta ,\psi
_{\theta })]]  \notag \\
 & =\arg \!\min_{\xi _{i}\in \Xi }\mathrm{E}_{\xi _{i}}[\mathrm{%
E}_{\theta_{i} |\xi _{i}}[\mathrm{E}_{\theta \backslash \theta
_{i}}[C(\theta ,\psi _{\mathrm{IBR}}^{\Theta |\xi _{i}})-C(\theta ,\psi
_{\theta })]]]  \notag \\
 & =\arg \!\min_{\xi _{i}\in \Xi }\mathrm{E}_{\theta _{i}}[\mathrm{%
E}_{\xi _{i}|\theta _{i}}[\mathrm{E}_{\theta \backslash \theta
_{i}}[C(\theta ,\psi _{\mathrm{IBR}}^{\Theta |\xi _{i}})-C(\theta ,\psi
_{\theta })]]]  \notag \\
& =\arg \!\min_{\xi _{i}\in \Xi }\mathrm{E}_{\theta _{i}}[\mathrm{E}_{\xi
_{i}|\theta _{i}}[\mathrm{E}_{\theta \backslash \theta _{i}}[C(\theta ,\psi
_{\mathrm{IBR}}^{\Theta |\xi _{i}})]]],  \label{eq:mocu-navid}
\end{align}
%\end{split}
%\label{eq:mocu-navid}
%\end{equation}%
where ``$\backslash$'' denotes set subtraction in the subscripts. The second line holds because only the posterior distribution of $%
\theta _{i}$ depends on experiment $\xi _{i}$; and the last equality follows from the independence of $C(\theta ,\psi
_{\theta })$ from $\xi _{i}$. The last line is exactly the policy derived in 
\cite{Navid} but there the policy derivation was based on adding the
objective-based cost of experimental error to the previous notion of
objective cost of uncertainty, whereas here we directly apply the generalized
formulation of MOCU as we have formulated in Section \ref{sec:GeneralMOCU}. %\vspace{-0.1in}
%Since $\mathrm{E}_{\xi _{i}}\mathrm{E}_{\theta |\xi _{i}}=\mathrm{E}_{\xi
%_{i}}\mathrm{E}_{\theta _{i}|\xi _{i}}\mathrm{E}_{\theta \backslash \theta
%_{i}}=\mathrm{E}_{\theta _{i}}\mathrm{E}_{\xi _{i}|\theta _{i}}\mathrm{E}%
%_{\theta \backslash \theta _{i}}$, where ``$\backslash$'' denotes set subtraction in the subscripts,
%the first equality holds because only the posterior distribution of $%
%\theta _{i}$ depends on experiment $\xi _{i}$. According to %
%\eqref{eq:mocu-experimentvalue} and \eqref{eq:mocu-policy}, the optimal
%experiment can be derived: %\begin{equation}
%%\begin{split}
%\begin{align}
% \xi^{\ast }&=\arg \!\min_{\xi _{i}\in \Xi }\mathrm{E}_{\theta _{i}}[\mathrm{%
%E}_{\xi _{i}|\theta _{i}}[\mathrm{E}_{\theta \backslash \theta
%_{i}}[C(\theta ,\psi _{\mathrm{IBR}}^{\Theta |\xi _{i}})-C(\theta ,\psi
%_{\theta })]]]  \notag \\
%& =\arg \!\min_{\xi _{i}\in \Xi }\mathrm{E}_{\theta _{i}}[\mathrm{E}_{\xi
%_{i}|\theta _{i}}[\mathrm{E}_{\theta \backslash \theta _{i}}[C(\theta ,\psi
%_{\mathrm{IBR}}^{\Theta |\xi _{i}})]]],  \label{eq:mocu-navid}
%\end{align}
%%\end{split}
%%\label{eq:mocu-navid}
%%\end{equation}%
%where the second equality follows from the independence of $C(\theta ,\psi
%_{\theta })$ from $\xi _{i}$.
\section{Connection of MOCU-based Experimental Design with KG and EGO}\label{sec:KG-EGO-derivation}

%\vspace{-0.1in}
Knowledge Gradient (KG) \cite{kg2008,correlatedKG}, which is used in
different fields, from drug discovery to material design \cite%
{kgMaterial1,kgMaterialnew}, was originally introduced as a solution to an
offline ranking and selection problem, where the assumption is that there
are $A\geq 2$ actions (alternatives) that can be selected, i.e., ${\Psi }%
=\{\psi _{1},\ldots,\psi _{A}\}$. Each action has an unknown true reward
(sign-flipped cost) and at each time step an experiment provides a noisy
observation of the reward of a selected action. There is a limited budget ($%
B $) of the number of measurements we can make before the time arrives to
decide which action is the best, that being the one having the lowest
expected cost (or the highest expected reward).

The assumption is that we have Gaussian prior beliefs over the unknown
rewards, either independent Gaussian beliefs over the rewards when the
rewards of different actions are uncorrelated, or a joint Gaussian belief
when the rewards are correlated. In the independent case, for each
action-reward pair $(\psi _{i},\theta _{\psi _{i}})$, $\theta _{\psi
_{i}}\sim N(m_{\psi _{i}},\beta _{\psi _{i}})$. In the correlated case, the
vector of rewards, $[\theta _{\psi _{1}},\ldots,\theta _{\psi _{A}}]$, has a
multivariate Gaussian distribution $N(m,\Sigma )$ with the mean vector $%
m=[m_{\psi _{1}},\ldots,m_{\psi _{A}}]$ and covariance matrix $\Sigma $,
with diagonal entries $[\beta _{\psi _{1}},\ldots,\beta _{\psi _{A}}]$. If
the selected action to be applied at $t$ is $\psi ^{t}$, then the observed
noisy reward of $\psi ^{t}$ at that iteration is $\xi ^{t}=\theta _{\psi
^{t}}+\epsilon ^{t}$, where $\theta _{\psi ^{t}}$ is unknown and $\epsilon
^{t}\sim N(0,\lambda _{\psi ^{t}})$ is independent of the reward of $\psi
^{t}$.

Here, the underlying system to learn is the unknown reward function and each
possible model is fully described by a reward vector $\theta =[\theta _{\psi
_{1}},\theta _{\psi _{2}},\ldots,\theta _{\psi _{A}}]$ in the uncertainty
class $\Theta $. For the independent case, $\pi (\theta
)=\prod_{i=1}^{A}N(m_{\psi _{i}},\beta _{\psi _{i}})$. For the correlated
case, $\pi (\theta )=N(m,\Sigma )$. The experiment space is $\Xi =\{\xi
_{1},\ldots,\xi _{A}\}$, where experiment $\xi _{i}$ corresponds to applying 
$\psi _{i}$ and getting a noisy observation of its reward $\theta _{\psi
_{i}}$, that is, measuring $\theta _{\psi _{i}}$ with observation noise,
where $\xi _{i}|\theta _{\psi _{i}}\sim N(\theta _{\psi _{i}},\lambda _{\psi
_{i}})$. In the independent case the state of knowledge at each time point $%
t $ is captured by the posterior values of the means and variances for the
rewards after incorporating observations $\xi ^{:t}$ as $S^{t}=[(m_{\psi
}^{t},\beta _{\psi }^{t})]_{\psi \in \Psi }$, and in the correlated case by
the posterior vector of means and a covariance matrix after observing $\xi
^{:t}$ as $S^{t}=(m^{t},\Sigma ^{t})$, where $m^{t}=[m_{\psi
_{1}}^{t},\ldots,m_{\psi _{A}}^{t}]$ and the diagonal of $\Sigma ^{t}$ is
the vector $[\beta _{\psi _{1}}^{t},\ldots,\beta _{\psi _{A}}^{t}]$. The
probability space $\Theta |\xi ^{:t}$ is equal to $\Theta |S^{t}$ and the
cost function is $C(\theta ,\psi )=-\theta _{\psi }$.

For this problem, the IBR action at time step $t$ is %\begin{equation}
%\begin{split}
\begin{align}
\psi _{\mathrm{IBR}}^{\Theta |\xi ^{:t}}=& \arg \!\min_{\psi \in \Psi }%
\mathrm{E}_{\Theta |\xi ^{:t}}\big[C(\theta ,\psi )\big]=\arg \!\min_{\psi
\in \Psi }\mathrm{E}_{\Theta |\xi ^{:t}}\big[-\theta _{\psi }\big]  \notag \\
=& \arg \!\max_{\psi \in \Psi }\mathrm{E}_{\Theta |\xi ^{:t}}\big[\theta
_{\psi }\big]=\arg \!\max_{\psi \in \Psi }m_{\psi }^{t},
\end{align}
%\end{split}%
%\end{equation}%
Again, by \eqref{eq:mocu-experimentvalue} and \eqref{eq:mocu-policy}, the optimal
experiment at time step $t$ can be derived:  
\begin{align}
\xi^{\ast ,t}& =\arg \!\min_{\xi _{i}\in \Xi }\mathrm{E}_{\mathcal{\xi }%
_{i}|\xi ^{:t}}[\mathrm{E}_{\theta |\xi _{i},\xi ^{:t}}[C(\theta ,\psi _{%
\mathrm{IBR}}^{\Theta |\xi ^{:t},\xi _{i}})]] \notag \\ & \quad \quad \quad \quad \quad \quad- \mathrm{E}_{\theta |\xi ^{:t}}[C(\theta ,\psi _{%
\mathrm{IBR}}^{\Theta |\xi ^{:t}})]  \notag \\
%& \quad \quad \quad \quad \mathrm{E}_{\theta |\xi ^{:t}}[C(\theta ,\psi _{\mathrm{IBR}}^{\Theta |\xi ^{:t}})]  \notag \\
& =\arg \!\min_{\xi _{i}\in \Xi }\mathrm{E}_{\xi _{i}|\xi ^{:t}}\Big[\mathrm{%
E}_{\theta |\xi ^{:t+1}}\big[-\theta _{\psi _{\mathrm{IBR}}^{\Theta |\xi
^{:t+1}}}\big]\Big] \notag \\ & \quad \quad \quad \quad \quad \quad - \mathrm{E}_{\theta |{\xi ^{:t}}}\big[-\theta
_{\psi _{\mathrm{IBR}}^{\Theta |\xi ^{:t}}}\big] \notag \\
%& \quad \quad \quad \quad \mathrm{E}_{\theta |{\xi ^{:t}}}\big[-\theta_{\psi _{\mathrm{IBR}}^{\Theta |\xi ^{:t}}}\big]  \notag \\
& =\arg \!\max_{\xi _{i}\in \Xi }\mathrm{E}_{\xi _{i}|\xi ^{:t}}\Big[\mathrm{%
E}_{\theta |\xi ^{:t+1}}\big[\theta _{\psi _{\mathrm{IBR}}^{\Theta |\xi
^{:t+1}}}\big]\Big]- \mathrm{E}_{\theta |{\xi ^{:t}}}\big[\theta _{\psi
_{\mathrm{IBR}}^{\Theta |\xi ^{:t}}}\big]  \notag \\
%& \quad \quad \quad \quad \mathrm{E}_{\theta |{\xi ^{:t}}}\big[\theta _{\psi_{\mathrm{IBR}}^{\Theta |\xi ^{:t}}}\big]  \notag \\
& =\arg \!\max_{\xi _{i}\in \Xi }\mathrm{E}_{\xi _{i}|\xi ^{:t}}\Big[%
\max_{\psi ^{\prime }\in \Psi }m_{\psi ^{\prime }}^{t+1}\Big]-\max_{\psi
^{\prime }\in \Psi }m_{\psi ^{\prime }}^{t}.  \label{eq:mocu-ranking}
\end{align}
%The design \cite{correlatedKG} and \cite{Wang2015NestedBatchModeLA}. 
The derived policy~\eqref{eq:mocu-ranking} by direct
application of the generalized MOCU is exactly the same as the original KG policy in \cite{kg2008}, \cite{correlatedKG}, and \cite{Wang2015NestedBatchModeLA}. 
As KG is shown to be optimal when the horizon is a single measurement and
asymptotically optimal (the number of measurements goes to infinity), the
same holds for the MOCU-based policy for this problem.%\vspace{-6pt}

Efficient Global Optimization (EGO) \cite{ego}, which is based on expected
improvement (EI), is widely used for black-box optimization and experimental
design. As shown in \cite{kgMaterialnew}, KG reduces to EGO when there is no
observation noise and choosing the best action at each time step is limited
to selecting from the set of actions whose rewards have been previously
observed; that is, at each time step if we want to make a final decision as
to the best action to be applied, it must be an action whose performance has
been previously observed from the first time step up to that time. Thus,
MOCU-based learning can also be reduced to EGO under its model assumptions.
We will show this directly.

Consider the ranking and selection problem with no noise in the
observations, so that $\epsilon ^{t}=0$ for all $t$. Each experiment $\xi
_{i}$ corresponds to applying $\psi _{i}$ and observing the true value of $%
\theta _{\psi _{i}}$. Moreover, the choice of the best action at each time
step is confined to the set of actions whose rewards have been previously
observed. Let $\Psi ^{t}$ denote this set: $\Psi ^{t}=\{\psi ^{t^{\prime
}}\}_{t^{\prime }=1,\ldots,t}$. The IBR action at time $t$ is 
\begin{equation}
\psi _{\mathrm{IBR}}^{\Theta |\xi ^{:t}}=\arg \!\min_{\psi \in \Psi ^{t}}%
\mathrm{E}_{\Theta |\xi ^{:t}}\big[-\theta _{\psi }\big]=\arg \!\max_{\psi
\in \Psi ^{t}}\theta _{\psi },  \label{eq:IBR_action}
\end{equation}%
where the last equality is due to the fact that the reward of an action
whose performance is already observed is known, since there is no
observation noise. Let $Z^{t}=\{\xi ^{t^{\prime }}\}_{t^{\prime
}=1,\ldots,t} $ denote the set of experiments performed up to the current
time $t$, where experiment $\xi ^{t^{\prime }}$ corresponds to $\psi
^{t^{\prime }}$ being applied at $t^{\prime }$ and its reward being
observed, in other words, measurement of $\theta _{\psi ^{t^{\prime }}}$ at $%
t^{\prime }$. Since there is no point in measuring an action's reward more
than once, the next experiment is selected from the set of remaining
experiments, so that the experiment space at time step $t$ is $\Xi ^{t}=\Xi
\backslash Z^{t}$. From \eqref{eq:mocu-experimentvalue}, %
\eqref{eq:mocu-policy}, and \eqref{eq:IBR_action}, the optimal experiment
selected at $t$ is %\begin{equation}
%\begin{split}
%& e^{\ast ,t}=\arg \!\min_{\xi _{i}\in \Xi ^{t}}\mathrm{E}_{\xi _{i}|\xi
%^{:t}}\Big[\mathrm{E}_{\theta |\xi ^{:t+1}}\big[-\theta _{\psi
%_{\mathrm{IBR}}^{\Theta |\xi ^{:t+1}}}\big]\Big]- \\
%& \quad \quad \quad \quad \mathrm{E}_{\theta |{\xi ^{:t}}}\big[-\theta
%_{\psi _{\mathrm{IBR}}^{\Theta |\xi ^{:t}}}\big] \\
%=& \arg \!\max_{\xi _{i}\in \Xi \backslash Z^{t}}\mathrm{E}_{\theta _{\psi
%_{i}}|\xi ^{:t}}\Big[\max \big(\theta _{\psi _{i}},\max_{\psi ^{\prime
%}\in \Psi ^{t}}\theta _{\psi ^{\prime }}\big)\Big]- \\
%& \quad \quad \quad \quad \max_{\psi ^{\prime}\in \Psi ^{t}}\theta _{\psi ^{\prime }} \\
%=& \arg \!\max_{\xi _{i}\in \Xi \backslash Z^{t}}\mathrm{E}_{\theta _{\psi
%_{i}}|\xi ^{:t}}\Big[\max \big(\theta _{\psi _{i}}-\max_{\psi ^{\prime
%}\in \Psi ^{t}}\theta _{\psi ^{\prime }},0\big)\Big],
%\end{split}
%\label{eq:mocu-ego}
%\end{equation}%
\begin{align}
\xi^{\ast ,t}&=\arg \!\min_{\xi _{i}\in \Xi ^{t}}\mathrm{E}_{\xi _{i}|\xi
^{:t}}\Big[\mathrm{E}_{\theta |\xi ^{:t+1}}\big[-\theta _{\psi _{\mathrm{IBR}%
}^{\Theta |\xi ^{:t+1}}}\big]\Big] \notag \\ & \quad \quad \quad \quad \quad \quad -  \mathrm{E}_{\theta |{\xi ^{:t}}}\big[-\theta
_{\psi _{\mathrm{IBR}}^{\Theta |\xi ^{:t}}}\big] \notag \\
%& \quad \quad \quad \quad \mathrm{E}_{\theta |{\xi ^{:t}}}\big[-\theta_{\psi _{\mathrm{IBR}}^{\Theta |\xi ^{:t}}}\big]  \notag \\
&= \arg \!\max_{\xi _{i}\in \Xi \backslash Z^{t}}\mathrm{E}_{\theta _{\psi
_{i}}|\xi ^{:t}}\Big[\max \big(\theta _{\psi _{i}},\max_{\psi ^{\prime }\in
\Psi ^{t}}\theta _{\psi ^{\prime }}\big)\Big] \notag \\ & \quad \quad \quad \quad \quad \quad - \max_{\psi ^{\prime}\in \Psi ^{t}}\theta _{\psi
^{\prime }}  \notag \\
%& \quad \quad \quad \quad \max_{\psi ^{\prime}\in \Psi ^{t}}\theta _{\psi^{\prime }}  \notag \\
&= \arg \!\max_{\xi _{i}\in \Xi \backslash Z^{t}}\mathrm{E}_{\theta _{\psi
_{i}}|\xi ^{:t}}\Big[\max \big(\theta _{\psi _{i}}-\max_{\psi ^{\prime }\in
\Psi ^{t}}\theta _{\psi ^{\prime }},0\big)\Big],  \label{eq:mocu-ego}
\end{align}
which is exactly the EGO policy in \cite{ego}.

There are fundamental differences between the general MOCU formulation and
KG (or EGO): (1) with MOCU the experiment space and action space can be
different, enabling more flexible experimental design compared to the
assumption of the same experiment and action space in KG (or EGO); (2) MOCU
considers the uncertainty directly on the underlying physical model, which
allows direct incorporation of prior knowledge regarding the underlying
system, whereas in KG (or EGO) the uncertainty is considered on the reward
function and there is no direct connection between prior assumptions and the
underlying physical model.

\section{A Simulation Study to Compare MOCU-based Experimental Design and KG}
\label{sec:Simulation}

In this section, we perform a simulation study to illustrate the flexibility of MOCU-based experimental design compared to KG, especially the importance of the flexibility of dissecting the uncertainty class assumptions to better incorporate prior knowledge regarding the underlying model. Here we compare the experimental design performances by MOCU and KG based on a simulated quadratic function example
with one input variable as the underlying reward
function that we want to maximize: $f(\theta ,\psi )=\theta _{1}\psi
^{2}+\theta _{2}\psi +\theta _{3}$, i.e. $C(\theta ,\psi )=-f(\theta ,\psi )$. The observation noise is additive Gaussian with the distribution $%
N(0,\theta _{4}^{2})$. In this simulation model, $\theta _{1}$, $\theta _{2}$, $\theta _{3}$ and $\theta _{4}$ are unknown parameters. We take $\Psi =\{\psi _{1},...,\psi _{20}\}=\{0.5,1,1.5,...,10\}$ as the set of actions (possible input
values $\psi$ ). The corresponding
experiment for each action is to apply $\psi _{i}$ so that we can observe the outcome $\xi _{i}$ (the reward):  
\begin{equation}
\xi _{i}|\theta \sim N(\theta _{1}\psi_i ^{2}+\theta _{2}\psi_i +\theta
_{3},\theta _{4}^{2}).
\end{equation}%
Note that as shown in Section \ref{sec:KG-EGO-derivation}, under model assumptions of KG,  MOCU-based experimental design results in the same policy as KG. But here, as opposed to KG that directly models the rewards (and corresponding costs) of actions with Gaussian distributions with (prior) fixed parameter values (either known or estimated), MOCU-based experimental design computes the generalized MOCU by modeling the uncertainty of the reward function by incorporating the uncertainty over the underlying parameters, to guide the experimental design procedure.

For both MOCU-based experimental design and KG, we assume that there is no prior knowledge on the model parameters $\mathbf{\theta} = [\theta_1, \theta_2, \theta_3, \theta_4]$. For MOCU, the non-informative prior $\pi (\theta ) \propto \theta _{4}^{-2}$ is used, which updates to a Gaussian-inverse-gamma distribution ($\pi^*(\theta)$) when measurements become available when experiments are carried out in sequence. For KG, to model the rewards of actions directly with correlated Gaussian distributions, approximate beliefs are constructed at each experiment since the noise variance is unknown and no joint
Gaussian prior distribution exists over the reward values of the actions. For this approximation, following \cite{kgGPR} and \cite%
{kgMaterialnew}, a Gaussian process regression (GPR) model \cite%
{rasmussen2006gaussian} with a quadratic basis (mean) function and a squared
exponential covariance matrix with additive Gaussian observation noise is
trained using the measurements performed (experiment outcomes observed) up
to that time step (by maximizing the marginal log-likelihood
of the observations).

In our simulation, $\theta _{1}$ is drawn from $%
U(-5,2)$ ($U(a,b)$ denotes the uniform distribution over the interval $(a,b)$); $%
\theta _{2}$ is set to $-2\theta _{1}r$, where $r$ is drawn from $U(-2.5,13)$%
; $\theta _{3}$ is sampled from $U(-5,5)$; and $\theta _{4}
$ is set to $\sigma (f) \times w$, where $w\sim U(0.075,0.7)$ and $\sigma (f)$
denotes the true standard deviation of the reward values of actions based on the given model parameters. Each simulation starts with four randomly selected actions, for which noisy observations of their rewards are simulated as initial training data to
both MOCU-based experimental design and KG. The sequential experimental design procedures based on MOCU and KG
are both continued for five iterations. For KG at each time step $t$, the
(posterior) vector of means ($m^{t}$), the covariance matrix ($\Sigma ^{t}$%
), and the noise variance are estimated by training a GPR model on the
available measurements, and the next experiment is selected by %
\eqref{eq:mocu-ranking}. For MOCU-based experimental design at each time step $t$, the
(posterior) Gaussian-inverse-gamma distribution after incorporating the
available measurements is used in \eqref{eq:mocu-policy2} to optimally
select the next experiment. 

To compare the performances, we check the
average opportunity cost metric, defined as the difference between the true
maximum of the reward among all the actions and the true reward of the action
selected as the best one based on two experimental design strategies. Note that this best action might be different from the next suggested experiment by each policy. The best action at each time step is the one that would be selected to be applied if the iterative experiments are stopped at that time. In other words, each experimental design policy suggests the next experiment, and after observing the outcome and based on its updated beliefs selects the best action (that would be applied if the iterative experiments were to stop) and the next experiment to be performed (if experimental budget is not exhausted). When following the MOCU-based policy, the next suggested experiment is the minimizer of the expected remaining MOCU, but the best action at each time step is the IBR action that maximizes (minimizes) the expectation of the reward (cost) with
respect to the (posterior) Gaussian-inverse-gamma distribution of uncertain
parameters based on the latest belief at that time step. When following the KG policy, the best action at each time step is the one that maximizes the
(posterior) GPR mean value at that time step which might be different from the suggested next experiment by KG.

\begin{figure}[tph]
%\begin{center}
\centering
\includegraphics[width=.99\linewidth]{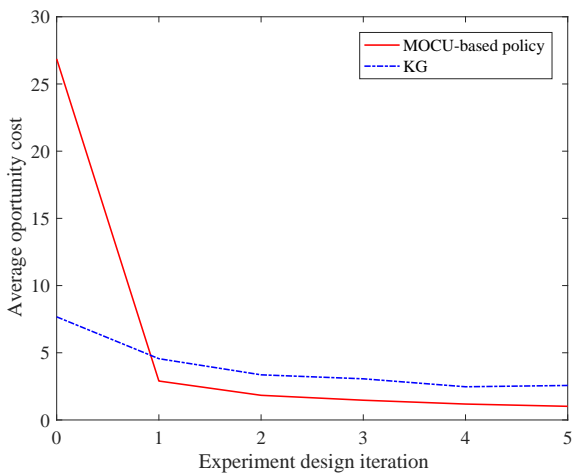}
\caption{Average opportunity cost of MOCU-based policy compared with KG
policy.}
\label{fig:comparison}
\end{figure}

Figure \ref{fig:comparison} illustrates the average opportunity cost for
MOCU-based experimental design and KG over 1,000 simulation runs. As can be seen from
the figure, as soon as the experimental design iterations begin MOCU-based
policy consistently has the lower average opportunity cost compared to KG. This confirms that directly incorporating the model uncertainty (the uncertainty of
model parameters in this simulation study as we assume that we have the model functional form) in the generalized MOCU framework results in a better experimental design policy. Note that at iteration 0 no experiment selection by any of the methods is performed, and only four randomly selected experiment outcomes are available. Since the flat (non-informative) prior is assumed for the parameters in the MOCU-based framework, the IBR action selection as the best action can be very conservative before beginning the experimental design procedure. The maximizer of the direct approximation of the reward function by GPR at iteration 0 is better than the IBR action for this simple simulation model. But as soon as the first experiment is selected by the policies, MOCU-based policy greatly reduces the uncertainty pertaining to the objective very sharply with the observed measurements and performs consistently better than KG.

\section{Conclusions}

In this letter, we present a generalized MOCU framework, leading to the MOCU-based experimental design pertaining to the maximum uncertainty reduction of differential cost with respect to the actual operational objectives. The proposed framework fits into Lindley's utility paradigm \cite{ldvbsar72sa} in classical Bayesian experimental design and is more flexible for the development of corresponding experimental design strategies for different real-world applications compared to the existing KG and EGO methods with their corresponding model assumptions. 
As we have shown in the simulation study (Section \ref{sec:Simulation}) and in the recent applications to life and materials science (Sections \ref{sec:Materials} and \ref{sec:Life}), our generalized MOCU framework, with the benefits from flexible dissection of the uncertainty class, action (operator) space, experiment space, and utility function depending on operational objectives, can lead to better objective-based uncertainty quantification and thereafter better experimental design to converge to desired objectives with smaller operational cost. 

\section*{Acknowledgments}

The authors acknowledge the support of NSF through the projects \emph{CAREER: Knowledge-driven Analytics, Model Uncertainty, and Experiment Design}, NSF-CCF-1553281; and \emph{DMREF: Accelerating the Development of Phase-Transforming Heterogeneous Materials: Application to High Temperature Shape Memory Alloys}, NSF-CMMI-1534534.

\bibliographystyle{./IEEEbib}
\bibliography{ref4}

\begin{thebibliography}{10}

\bibitem{mocu}
Byung-Jun Yoon, Xiaoning Qian, and Edward~R. Dougherty,
\newblock ``Quantifying the objective cost of uncertainty in complex dynamical
  systems,''
\newblock {\em IEEE Transactions on Signal Processing}, vol. 61, no. 9, pp.
  2256--2266, May 2013.

\bibitem{Dehghannasirimocu}
Roozbeh Dehghannasiri, Byung-Jun Yoon, and Edward~R. Dougherty,
\newblock ``Optimal experimental design for gene regulatory networks in the
  presence of uncertainty,''
\newblock {\em IEEE/ACM Trans. Comput. Biol. Bioinformatics}, vol. 12, no. 4,
  pp. 938--950, July 2015.

\bibitem{roozbehCanonical}
Roozbeh Dehghannasiri, Xiaoning Qian, and Edward~R. Dougherty,
\newblock ``Optimal experimental design in the context of canonical
  expansions,''
\newblock {\em IET Signal Processing}, vol. 11, pp. 942--951, October 2017.

\bibitem{ldvbsar72sa}
Dennis~V. Lindley,
\newblock {\em {B}ayesian {S}tatistics, A Review},
\newblock SIAM, Philadelphia, 1972.

\bibitem{kg2008}
Peter~I. Frazier, Warren~B. Powell, and Savas Dayanik,
\newblock ``A knowledge-gradient policy for sequential information
  collection,''
\newblock {\em SIAM Journal on Control and Optimization}, vol. 47, no. 5, pp.
  2410--2439, 2008.

\bibitem{correlatedKG}
Peter~I. Frazier, Warren~B. Powell, and Savas Dayanik,
\newblock ``The knowledge-gradient policy for correlated normal beliefs,''
\newblock {\em INFORMS Journal on Computing}, vol. 21, no. 4, pp. 599--613,
  2009.

\bibitem{ego}
Donald~R Jones, Matthias Schonlau, and William~J Welch,
\newblock ``Efficient global optimization of expensive black-box functions,''
\newblock {\em Journal of Global optimization}, vol. 13, no. 4, pp. 455--492,
  1998.

\bibitem{Imani_bioinf}
Mahdi Imani, Roozbeh Dehghannasiri, Ulisses~M. Braga-Neto, and Edward~R.
  Dougherty,
\newblock ``{MOCU} significantly outperforms entropy for experimental design,''
\newblock {\em Submitted}, 2017.

\bibitem{Navid}
Daniel~N. Mohsenizadeh, Roozbeh Dehghannasiri, and Edward~R. Dougherty,
\newblock ``Optimal objective-based experimental design for uncertain dynamical
  gene networks with experimental error,''
\newblock {\em IEEE/ACM Transactions on Computational Biology and
  Bioinformatics}, 2017,
\newblock doi: 10.1109/TCBB.2016.2602873.

\bibitem{DALTON1}
Lori~A. Dalton and Edward~R. Dougherty,
\newblock ``Optimal classifiers with minimum expected error within a {B}ayesian
  framework--part i: Discrete and {G}aussian models,''
\newblock {\em Pattern Recognition}, vol. 46, no. 5, pp. 1301--1314, 2013.

\bibitem{DALTON2}
Lori~A. Dalton and Edward~R. Dougherty,
\newblock ``Intrinsically optimal {B}ayesian robust filtering,''
\newblock {\em IEEE Transactions on Signal Processing}, vol. 62, no. 3, pp.
  657--670, Feb 2014.

\bibitem{RoozbehKalman}
Roozbeh Dehghannasiri, Mohammad~S. Esfahani, and Edward~R. Dougherty,
\newblock ``Intrinsically {B}ayesian robust {K}alman filter: An innovation
  process approach,''
\newblock {\em IEEE Transactions on Signal Processing}, vol. 65, no. 10, pp.
  2531--2546, May 2017.

\bibitem{s59core59as}
M.~Stone,
\newblock ``Application of a measure of information to the design and
  comparison of regression experiments,''
\newblock {\em Annals of Mathematical Statistics}, vol. 30, pp. 55--70, 1959.

\bibitem{gh63}
Morris~H. DeGroot,
\newblock ``Uncertainty, information and sequential experiments,''
\newblock {\em Annals of Mathematical Statistics}, vol. 33, no. 2, pp.
  404--419, 1962.

\bibitem{DeGroot1986}
Morris~H. DeGroot,
\newblock {\em Concepts of Information Based on Utility}, pp. 265--275,
\newblock Springer Netherlands, Dordrecht, 1986.

\bibitem{b79}
Jos{\'e}~M. Bernardo,
\newblock ``Expected information as expected utility,''
\newblock {\em Annals of Statistics}, vol. 7, no. 3, pp. 686--690, 1979.

\bibitem{cv95}
Kathryn Chaloner and Isabella Verdinelli,
\newblock ``{B}ayesian experimental design: {A} review,''
\newblock {\em Statistical Science}, vol. 10, no. 3, pp. 273--304, 1995.

\bibitem{Qian2016}
Xiaoning Qian and Edward~R. Dougherty,
\newblock ``Bayesian regression with network prior: {Optimal B}ayesian
  filtering perspective,''
\newblock {\em IEEE Transactions on Signal Processing}, vol. 64, no. 23, pp.
  6243--6253, 2016.

\bibitem{RoozbehMaterial}
Roozbeh Dehghannasiri, Dezhen Xue, Prasanna~V. Balachandran, Mohammadmahdi~R.
  Yousefi, Lori~A. Dalton, Turab Lookman, and Edward~R. Dougherty,
\newblock ``Optimal experimental design for materials discovery,''
\newblock {\em Computational Materials Science}, vol. 129, no. Supplement C,
  pp. 311--322, 2017.

\bibitem{Navid1}
Daniel~N. Mohsenizadeh, Jianping Hua, Michael Bittner, and Edward~R. Dougherty,
\newblock ``Dynamical modeling of uncertain interaction-based genomic
  networks,''
\newblock {\em BMC Bioinformatics}, vol. 16, no. 13, pp. S3, Dec 2015.

\bibitem{kgMaterial1}
Si~Chen, Kristofer-Roy~G. Reyes, Maneesh~K. Gupta, Michael~C. McAlpine, and
  Warren~B. Powell,
\newblock ``Optimal learning in experimental design using the knowledge
  gradient policy with application to characterizing nanoemulsion stability,''
\newblock {\em SIAM/ASA Journal on Uncertainty Quantification}, vol. 3, no. 1,
  pp. 320--345, 2015.

\bibitem{kgMaterialnew}
Peter~I. Frazier and Jialei Wang,
\newblock ``Bayesian optimization for materials design,''
\newblock {\em Information science for materials discovery and design}, vol.
  225, pp. 45--57, 2016.

\bibitem{Wang2015NestedBatchModeLA}
Yingfei Wang, Kristofer~G. Reyes, Keith~A. Brown, Chad~A. Mirkin, and Warren~B.
  Powell,
\newblock ``Nested-batch-mode learning and stochastic optimization with an
  application to sequential multistage testing in materials science,''
\newblock {\em SIAM J. Scientific Computing}, vol. 37, 2015.

\bibitem{kgGPR}
Warren Scott, Peter Frazier, and Warren Powell,
\newblock ``The correlated knowledge gradient for simulation optimization of
  continuous parameters using {G}aussian process regression,''
\newblock {\em SIAM Journal on Optimization}, vol. 21, no. 3, pp. 996--1026,
  2011.

\bibitem{rasmussen2006gaussian}
Carl~E. Rasmussen and Christopher~K.I. Williams,
\newblock {\em Gaussian Processes for Machine Learning},
\newblock Adaptative computation and machine learning series. University Press
  Group Limited, 2006.

\end{thebibliography}

\end{document}